\documentclass[runningheads,a4paper,11pt]{llncs}
\usepackage[english]{babel}

\usepackage[dvips]{graphicx}
\usepackage{fancyhdr}
\usepackage{hyperref}
\usepackage[usenames,dvipsnames]{pstricks}
\usepackage{epsfig}
\usepackage{subfloat}
\usepackage{graphicx}
\usepackage{hyperref} 
\usepackage{url}

\usepackage{cite}
\usepackage{amsfonts}

\usepackage[dvips]{graphicx}
\usepackage[usenames,dvipsnames]{pstricks}
\usepackage{epsfig}
\usepackage{subfigure}
\usepackage{pst-grad} 
\usepackage{url}

\usepackage{amsmath}
\usepackage{amssymb}
\usepackage{pifont}
\usepackage{hyperref}
\usepackage{pbox}

\newcommand{\xmark}{\ding{55}}%

\usepackage[dvips]{graphicx}

\usepackage[usenames,dvipsnames]{pstricks}
\usepackage{epsfig}
\usepackage{subfigure}
\usepackage{pst-grad} 

\usepackage{url}
\usepackage[lined, boxruled, linesnumbered]{algorithm2e}

\usepackage{amstext}
\usepackage{array}
\usepackage{url}
\usepackage{mathtools}
\usepackage{color}

\usepackage{enumerate}
\usepackage{multicol}
\usepackage{setspace}

\usepackage{algorithmic} 

\normalfont

\begin{document}
\title{Still Wrong Use of Pairings in Cryptography}
\author{Mehmet Sab{\i}r Kiraz and Osmanbey Uzunkol\\ Mathematical and Computational Sciences Labs\\ T\"{U}B\.{I}TAK B\.{I}LGEM, Turkey\\
\{mehmet.kiraz,osmanbey.uzunkol\}@tubitak.gov.tr }

\institute{}

\titlerunning{Still Wrong Use of Pairings in Cryptography}

\maketitle

% The paper headers

\begin{abstract}
%\boldmath

Several pairing-based cryptographic protocols are recently proposed with a wide variety of new novel applications including the ones in emerging technologies like cloud computing, internet of things (IoT), e-health systems and wearable technologies. There have been however a wide range of incorrect use of these primitives. The paper of Galbraith, Paterson, and Smart (2006) pointed out most of the issues related to the incorrect use of pairing-based cryptography. However, we noticed that some recently proposed applications still do not use these primitives correctly. This leads to unrealizable, insecure or too inefficient designs of pairing-based protocols. We observed that one reason is not being aware of the recent advancements on solving the discrete logarithm problems in some groups. The main purpose of this article is to give an understandable, informative, and the most up-to-date criteria for the correct use of pairing-based cryptography. We thereby deliberately avoid most of the technical details and rather give special emphasis on the importance of the correct use of bilinear maps by realizing secure cryptographic protocols. We list a collection of some recent papers having wrong security assumptions or realizability/efficiency issues. Finally, we give a compact and an up-to-date recipe of the correct use of pairings.


\end{abstract}

% make the title area

\section{Introduction}\label{sc:intro}

Pairing-based cryptography has received much attention because of wide variety of its immediately deployable applications. These applications include identity-based encryption, functional and attribute-based encryption, searchable encryption, short/group/ring signatures, signcryption, homomorphic linear authenticators for integrity checking, security, privacy and integrity solutions for cloud computing and Internet of Things (IoT), e-health systems, and wearable technologies. We refer to Appendix \ref{appendix} for a selected list of some novel applications using pairing-based cryptography. In practice, Voltage Security (now an HP company) and Trend Micro are the most well-known companies utilizing the pairing-based security solutions \cite{Nist15}.

There have been unfortunately a collection of results using the pairing-based primitives incorrectly. In fact, Galbraith, Paterson, and Smart drew attention to the potential problems related to using these ``black boxes'' incorrectly \cite{GPS08}. However, we notice that many recent research papers still have security vulnerabilities, realizability issues and/or efficiency problems since the appearance of \cite{GPS08}. These papers (surprisingly) either have pairing related wrong security assumptions regarding the infeasibility of certain computational problems and/or efficiency issues. The main reason is to use pairing-based primitives as ``black-boxes'' without giving attention on the concrete realizations of these primitives.

The security of pairing-based cryptosystems relies on the difficulty of various computationally hard problems related to the discrete logarithm problem (DLP). However, there are also new attacks on the DLP on some groups \cite{GGMZ13, joux:2014a,rodriguez:2014a,menezes:2015a,granger:2014a}. Furthermore, very recent results on solving the DLP for finite fields of medium characteristics and composite degrees size have also significant consequences on the choice of primitives for pairing based cryptography \cite{Kim2016,Sarkar2016,Jeong2016}. In particular, designers need to update the previously used key sizes for composite embedding degrees. Both attacks have also major consequences on the design of secure cryptographic protocols based on pairing-based cryptography. Hence, one additional relatively new reason of incorrect use of pairing-based primitives is to ignore these recent technical advancements in solving the DLP which make certain security assumptions incorrect. The complexity of these mathematical preliminaries is undoubtedly the reason of neglecting the realization concerns in the design of pairing-based protocols.

In this work, our aim is to highlight the importance of correct choices and their affect on the realizability of abstract pairing requirements to design cryptographic protocols with prescribed level of security, realizability and desired efficiency. In this respect, we briefly survey the most recent attacks against pairing-based cryptography which have direct effects on the designs and the security models of cryptographic protocols. We further emphasize the concerns in the paper of Galbraith, Paterson, and Smart \cite{GPS08} together with further new security issues and their implications. We stress thereby that this paper does not propose new improvements or new mathematical techniques but deliberately give attention to the incorrect use of pairings. Therefore, our purpose is to give an informative and less technical overview of pairing-based mechanisms which briefly surveys the various pairing-based real-world applications while taking the recent mathematical improvements having direct security and efficiency implications into account. Furthermore, we give a more detailed list on various pairing related hard problems together with their relation to the security assumptions of the underlying pairing-based protocols. We finally propose a compact and state-of-the-art recipe for designers to take it into consideration for proper usage. This recipe covers main security and realizability issues of pairing-based cryptography which may help the designers to use the primitives correctly.

\section{Basics for Pairing-Based Cryptography}\label{prel}

We begin with the abstract pairing requirements and different types of bilinear maps used in cryptographic protocols.

Let $(\mathbf{G}_1, + )$ and $(\mathbf{G}_2, +)$ be two additive cyclic groups of (nearly) prime order $q$ with $\mathbf{G}_1= <P>$ and $\mathbf{G}_2= <Q>$, $(\mathbf{G}_T, \cdot)$ be a multiplicative cyclic group of order $q$ with $\mathbf{G}_T= <g>$. We write as usual $0$ for the identity elements of $\mathbf{G}_1$, $\mathbf{G}_2$ and $1$ for $\mathbf{G}_T$. A \textit{pairing} or a \textit{bilinear map} is a map $e:\ \mathbf{G}_1\times \mathbf{G}_2\rightarrow \mathbf{G}_T$ satisfying the following properties:

\begin{itemize}
\item{\textbf{Bilinearity:}} For all $P_1,P'_1 \in \mathbf{G}_1, Q_1,Q'_1 \in \mathbf{G}_2$, $e$ is a group homomorphism in each component, i.e. 
\begin{enumerate}
\item $e(P_1+P'_1,Q_1)=e(P_1,Q_1)\cdot e(P'_1,Q_1)$,
\item $e(P_1,Q_1+Q'_1)=e(P_1,Q_1)\cdot e(P_1,Q'_1)$.
\end{enumerate}
\item{\textbf{Non-degeneracy:}} $e$ is non-degenerate in each component, i.e.
\begin{enumerate}
\item For all $P_1\in \mathbf{G}_1$, $P_1\neq 0$, there is an element $Q_1 \in \mathbf{G}_2$ such that $e(P_1,Q_1)\neq 1$,
\item For all $Q_1\in \mathbf{G}_2$, $Q_1\neq 0$, there is an element $P_1 \in \mathbf{G}_1$ such that $e(P_1,Q_1)\neq 1$.
\end{enumerate}
\item{\textbf{Computability:}} There exists an algorithm which computes the bilinear map $e$ efficiently.
\end{itemize}

There are essentially 4 types of bilinear maps \cite{GPS08,Sha05} used in the design of pairing-based protocols depending on the special requirements such as short representation, hashing to a group element, efficient homomorphisms.

\begin{itemize}
\item \textbf{Type-1:} $\mathbf{G}_1=\mathbf{G}_2$. In this case there exists no short representations for the elements of $\mathbf{G}_1$. 
\item \textbf{Type-2:} $\mathbf{G}_1\neq \mathbf{G}_2$ and there is an efficiently computable homomorphism $\phi:\mathbf{G}_2 \rightarrow \mathbf{G}_1$. In this case no efficient secure hashing to the elements in $\mathbf{G}_2$ is possible.
\item \textbf{Type-3:} $\mathbf{G}_1\neq \mathbf{G}_2$ and there exists no efficiently computable homomorphism $\phi:\mathbf{G}_2\rightarrow \mathbf{G}_1$. 

\item \textbf{Type-4:} $\mathbf{G}_1\neq \mathbf{G}_2$ and there exists an efficiently computable homomorphism $\phi:\mathbf{G}_2\rightarrow \mathbf{G}_1$ as in the case of the Type-2 setting but with an efficient secure hashing method to a group element \cite{Sha05}. Security proofs can be quite cumbersome in this setting as discussed in \cite{CCS07}. We note that this type is not generally used in protocol designs due to its inefficiency.
\end{itemize}

The main disadvantage of the Type-2 pairing is that there exists no random sampling algorithm from $\mathbf{G}_2$ (yielding to a secure hash function) which maps arbitrary elements to $\mathbf{G}_2$, \cite[pp. 3119]{GPS08}. Note that there exists a natural, efficient, and secure transformation of protocols using the Type-2 pairing into protocols using the Type-3 pairing \cite[Section 5]{CM11}.

The Type-1 setting is commonly called \textit{symmetric pairing} while other types are called \textit{asymmetric pairing}.
  
\subsubsection{Properties and Conversion of Types.} Since the situation $\mathbf{G}_1 \neq \mathbf{G}_2$ with efficiently computable homomorphisms (in both directions) is essentially the same with the Type-1 setting (by identifying the groups via explicit homomorphisms), we do not consider it separately. 

The main technical part of pairing-based cryptography is the pairing functions including Weil, Tate and Ate pairing defined mostly on the product of certain subgroups of low dimensional abelian varieties over finite fields (in practice either on subgroups of elliptic curves or jacobians of genus two hyperelliptic curves) \cite{BSS05}. 

Due to efficiency and realizability concerns of pairing-based protocols many ad hoc and conceptual conversion methods from one type of pairing to another one has been proposed \cite{Wat09,RCS12,CLLWW13}. Abe et al. \cite{AGOT14} proposed a generic framework converting not only the protocols with the Type-1 bilinear maps into the Type-3 setting but also converting corresponding security proofs using black-box reduction methods in the random oracle model. Akinyele et al. \cite{AGH15} have very recently given some concerns about the practicability of the elegant theoretic solution of \cite{AGOT14} and proposed an automated software tool transforming schemes using the Type-1 bilinear maps into the Type-3 setting. We note however that the proposed automated tool in \cite{AGH15} and generic frameworks in \cite{AGOT14} suffer from being inefficient when compared to their manual counterparts like \cite{Wat09,RCS12,CLLWW13}. In \cite[p. 20]{AGH15}, it is left as an open problem to generalize and systematize the manual advancement more efficiently for automated tools.

\subsubsection{Basic Computational Problems related to Pairing:}

For completeness of the paper, we briefly summarize the basic computational problems. Let $\mathbf{G}$ be a finite cyclic group of order $n$ and $P \in \mathbf{G}$ be its generator (here additively written). In order to use $\mathbf{G}$ for cryptographic purposes, we need the existence of the efficient algorithms available to compute in the group $\mathbf{G}$. Hence, the isomorphism between $(\mathbb{Z}/n\mathbb{Z},+)$ and $(\mathbf{G},+)$ can explicitly be given and efficiently computable via 

$$\phi:\ \mathbb{Z}/n\mathbb{Z}\mathbb\longrightarrow \mathbf{G}, \ a\mapsto aP.$$

The discrete logarithm problem (DLP) asks to find the preimage in $\mathbb{Z}/n\mathbb{Z}$ of an arbitrarily chosen element in $\mathbf{G}$, i.e. to find $a$ for a given pair $(P, aP)$, where $a$ is chosen randomly in $\mathbb{Z}/n\mathbb{Z}$. If the DLP is not intractable, in other words, if one can efficiently compute $\phi^{-1}(Q)$ for any $Q\in \mathbf{G}$, then all pairing related hardness assumptions will be wrong, i.e. they cannot be used to design secure cryptographic protocols.

The computational Diffie-Hellman problem is to compute $abP$ for a given triple $(P, aP, bP)$ where $a, b$ are chosen randomly in $\mathbb{Z}/n\mathbb{Z}$. The decisional Diffie-Hellman problem is to decide $Q \stackrel{?}{=} abP$ for a given quadruple $(P, aP, bP, Q)$ where $a, b$ are chosen randomly in $\mathbb{Z}/n\mathbb{Z}$. Due to Pohling-Hellman reduction it is usual to assume that $\mathbf{G}$ has a (nearly) prime order $r$ (or has a large prime order subgroup of order $r$, respectively). 

Provided that there exists a DLP solver for the image group $\mathbf{G}_T$ one has the following well-known fact by Frey-R\"uck and Menezes et al.:

\begin{theorem}\cite{Freyruckattack94,MOV93}\label{dlp-mov}
If there exists a bilinear map $e: \mathbf{G}_1 \times \mathbf{G}_2 \rightarrow \mathbf{G}_T$, then the DLP in $\mathbf{G}_1$ and $\mathbf{G}_2$ can be solved in polynomial time in the number of digits if there exists a DLP oracle for $\mathbf{G}_T$.
\end{theorem}

In the literature this attack is known as the MOV reduction attack.The result follows in a straightforward manner by the assumption that the pairing function $e$ is efficiently computable: Given $P \in \mathbf{G}_1$ and $aP \in \mathbf{G}_2$, we can compute $e(P, Q)$ $\in$ $\mathbf{G}_T$ and $e(P, aQ)$ = $e(P, Q)^a$ $\in$ $\mathbf{G}_T$. Then the DLP solver for $\mathbf{G}_T$ can be used to obtain $a$ \cite{Freyruckattack94,MOV93}. 

\section{Attacks on Pairing-Based Cryptography}

Several attacks on the DLP have recently been proposed improving the function field sieve algorithm in the multiplicative group of finite fields of small characteristics \cite{GGMZ13, joux:2014a,rodriguez:2014a,menezes:2015a,granger:2014a}. There are serious implications of these attacks on the security of pairing-based cryptography. More concretely, the use of symmetric pairing, and hence the use of pairing-friendly elliptic/hyperelliptic curves over fields of small characteristic are essentially useless \cite{granger:2014a,rodriguez:2014a}. Concrete attacks are performed for certain supersingular elliptic/hyperelliptic curves over $\mathbb{F}_2$ and $\mathbb{F}_3$, see \cite{rodriguez:2014a,granger:2014a}. Difficulties of generalizing these attacks on the elliptic curve setting are pointed out in a recent work of Massierer \cite{massierer:2015}. However, it can be argued more generally that the use of elliptic curves over finite fields of small characteristics in group-based cryptography has severe potential security threads. Especially, a very recent conjectural algorithm of Semaev \cite{semaev:2015a} shows that the believed security level of $285$ bits for a NIST elliptic curve over $\mathbb{F}_2^{571}$ can be reduced asymptotically to a security level of $101.7$ bits using a variant of Weil descent attack although there is not a consensus on the validity of such asymptotical conjectures \cite{GG:2015}.

In this section we summarize the attacks on pairing-based cryptography together with their implications.

\subsection{Recent Advances in solving the DLP}\label{DLPattacks}

%\subsection{Quasi-polynomial Attacks on the DLP and Bilinear Maps}\label{DLPattacks}

The difficulty of the DLP depends on the description of the underlying group $\mathbf{G}$. Indeed, Shoup showed that the intractability of the DLP is closely related to the algorithms available to the description of $\mathbf{G}$. He further shows that in generic groups the computation of the discrete logarithms costs at least $\Omega(\sqrt{p})$, where $p$ is the largest prime divisor of the order of $\mathbf{G}$ \cite{Shoup97}. In particular, computing the discrete logarithms in generic groups requires approximately $\sqrt{p}$ computation in $\mathbf{G}$. Recent advances on solving the DLP for finite fields have major consequences on the secure design of pairing based protocols. There are mainly two new attacks (1) quasi-polynomial DLP solver for small characteristics field, and (2) a more efficient variant of a DLP solver for medium characteristics fields of composite degree.

\begin{enumerate}
\item It was a well-known fact that the DLP for the multiplicative subgroups of finite fields is not as difficult as in the generic groups. However, many research has been done using these groups for cryptographic purposes by neglecting possible use of the algorithmic description of these groups to solve the discrete logarithm instances. The situation has been dramatically changed with the recent advancements of Joux et al. \cite{joux:2014a} and G\"ologlu et al. \cite{GGMZ13}. Recently, Granger et al. \cite{GKZ15} improved the result of Joux et al. \cite{joux:2014a} by proposing a new expected quasi-polynomial algorithm for solving the DLP for finite fields $\mathbf{F}_q^k$ with roughly $q\approx k$. These attacks removed the DLP for multiplicative subgroups of small characteristic finite fields from the list of intractable problems.

\item Very recent results on a variant of the number field sieve algorithm for solving the discrete logarithm problem in medium characteristics finite fields of composite degrees have direct consequences on the choice of key sizes for pairing based algorithms \cite{Kim2016,Sarkar2016,Jeong2016}. The complexity analysis of these new techniques suggests that doubling the sizes of the underlying elliptic curves is a conservative choice of maintaining the desired security level.

\end{enumerate}

Explicit realization of the bilinear maps can be done if $\mathbf{G}_T$ is a subgroup of the multiplicative group of a finite field. In particular, the DLP on $\mathbf{G}_1$ and $\mathbf{G}_2$ can be transferred into the subgroup of a finite field by Theorem \ref{dlp-mov}. Hence, the algorithms for solving the DLP for finite fields are applicable on the discrete logarithm instances of pairing groups $\mathbf{G}_1$ and $\mathbf{G}_2$. Therefore, these new attacks have direct consequences on the security of many pairing-based cryptographic applications if the characteristic of the field defining $\mathbf{G}_1$ is small \cite{JP15}. In fact, subsequent results applying the idea of this algorithm (combined with Frey-R\"uck and MOV attacks \cite{ADLNV06}) showed the fatal security issues for cryptographic protocols using the Type-1 bilinear maps \cite{joux:2014a,rodriguez:2014a,menezes:2015a,granger:2014a}.

In particular, for a group of size $n$ with $$L_n(\alpha,c)=\exp((c+o(1))(\log n)^\alpha (\log\log n)^{1-\alpha}),$$
where $0<\alpha<1, \ c>0$, Barbulescu et al. \cite{joux:2014a} improved the previous bound for solving the DLP of Joux from $L_n(1/4)$ for a specified $c$ to $n^{O(\log n)}$ for fields of the form $\mathbf{F}_q^k$ with roughly $q\approx k$. The key idea is to use a new and elegant approach for the descent phase. 

\subsubsection{Realization of Bilinear Maps: }In order to understand the impact of these attacks on the design of pairing-based cryptographic protocols, we now briefly summarize the realization of bilinear maps using suitable elliptic curves for cryptographic purposes.

Over a finite field $\mathbb{F}_q$ with $q=p^m$, $p$ a prime and $m\in\mathbb{N}$, the candidate groups $\mathbf{G}_1$ and $\mathbf{G}_2$ in the definition of bilinear maps are certain subgroups of a carefully chosen elliptic curve $E$ over $\mathbb{F}_q$. In particular, $\mathbf{G}_1$ is the $r-$th torsion subgroup $E(\mathbb{F}_q)[r]$ and $\mathbf{G}_2$ is a certain group related to the explicit realization of the bilinear map. We refer to \cite{BSS05} for further details.

The abstract condition on the efficient computation of 
$$e:\ \mathbf{G}_1\times \mathbf{G}_2\rightarrow \mathbf{G}_T$$
is realizable using Tate, Weil, Ate, and optimal pairing of elliptic curves, see for example \cite{FHess:2008a}. 
More concretely, given an elliptic curve $E$ over $\mathbb{F}_q$ the function $e$ takes rational points of $E$ over $\mathbb{F}_q$ or $\mathbb{F}_{q^k}$ as inputs and outputs elements of $\mathbb{F}_{q^k}^{*}$, where $k$ is the smallest integer with the property that $r$ divides $q^k-1$. the value $k$ is called the embedding degree of $E$ with respect to $r$. To achieve the desired security and efficiency in $\mathbf{G}_1$, $\mathbf{G}_2$ and $\mathbf{G}_T$ the ratio $\log q^k/\log r=k\rho$ with $\rho=\log q/\log r$ has to be balanced. We refer to \cite{FST10} for the details. For implementation and comparison purposes, one can consult Table \ref{recommendedsecuritylevels} following the lines of \cite{MGI09}.

\begin{table*}[htpb]
  \caption{Comparison with the recommended security levels for pairing groups, corresponding embedding degrees, $\rho$, and \#of modular multiplications (MM) over the prime field $\mathbb{F}_r$ for Ate and twisted Ate pairing using Barreto-Naehrig curves, (groups $\mathbf{G}_1$ and $\mathbf{G}_2$ have the same prime order $r$, $\mathbf{G}_T$ is a subgroup of $\mathbb{F}_{q^k}$ of order $r$, $k$ is the embedding degree which is the smallest integer such that $r|(q^k-1)$, i.e. $k$ is the order of $q \mod r$) \cite{MGI09}.}\label{recommendedsecuritylevels}

\begin{center}
\begin{tabular}{|c|c|c|c|c|c|}
    \hline
     Security& $r$ & $q^k$  &  $k$ with  & Ate & Twisted Ate\\ 
		Level (bits)&   (bits)&    (bits)&   $\rho$ $\approx$ 1 & Pairing ($k=12$) & Pairing ($k=12$)\\ \hline
   80 & 160   & 960-1280      &		6-8				 & 4647 MM & 7800 MM\\ \hline
    128  & 256 & 3000-5000 		 &		12-20		& 7119 MM & 12480 MM\\ \hline
		192  & 384 & 7000-9000 		 &		18-24		& 17007 MM & 31200 MM\\ \hline
		256 & 512 & 14000-18000		 &		28-36	& 33486 MM & 62400 MM \\ \hline
    \end{tabular}
\end{center}
\end{table*}

The probability that a randomly chosen (nearly) prime order elliptic curve $E$ has small enough embedding degree is negligibly small (generically $k$ is in $O(q)$) \cite{LMS04,FST10}. Hence, special pairing-friendly curves have to be constructed in order to realize an efficiently computable function $e$ with the property that the DLP is still intractable. Supersingular elliptic curves were initially the natural candidates of realizing such efficiently computable functions $e$ with the desired security level. The reason was that supersingular elliptic curves have embedding degrees $k=2,4$ or $6$ depending on whether $\mbox{char}(\mathbb{F}_q)\neq 2,3, \mbox{char}(\mathbb{F}_q)=2$ and $\mbox{char}(\mathbb{F}_q)=3$, respectively \cite{BSS05}. We refer to \cite{FST10} for further details on constructing curves with larger embedding degrees, i.e. the construction of ordinary pairing-friendly curves over prime fields with complex multiplication (CM) techniques (both families and individual curve constructions).

\subsubsection{Consequences of the Quasi-Polynomial Attacks on Pairing-Based Cryptography} As briefly outlined above, the new attacks for solving the DLP on the multiplicative subgroup of small characteristic finite fields have also dramatic consequences for the design of pairing-based protocols. In fact, these attacks showed either the insecurity of the use of supersingular elliptic curves (all pairing-friendly elliptic curves over fields of characteristic $2$ or $3$) or the inefficiency of their usage (all supersingular curves defined over a large characteristic prime field) in the pairing-based settings \cite{GGMZ13, joux:2014a,rodriguez:2014a,menezes:2015a,granger:2014a}. Since the Type-1 pairing can only be realized using supersingular elliptic/hyperelliptic curves \cite{GPS08} we see in Section \ref{hard_problems} that all Type-1 bilinear maps and the related protocols are either useless or regarded completely as insecure.

\subsubsection{Consequences of the Attacks for Composite Degree Finite Fields on Pairing-Based Cryptography} The attack of Barbulescu and Kim \cite{Kim2016} reduces the complexity of solving the DLP problem on $\mathbf{F}_{p}^n$ from $L_n(1/3,\sqrt[3]{96/9})$ to $L_n(1/3,\sqrt[3]{48/9})$ if $n=\kappa\eta$ with $\gcd(\kappa,\eta)=1$ and $\kappa,\eta>1$. Recently, Jeong and Kim removed the condition $\gcd(\kappa,\eta)=1$ implying that if $n$ is composite, the previous key sizes could be doubled asymptotically to guarantee the same security level of the disrete logorithm problem. Since, most pairing friendly elliptic curves have composite embedding degree (i.e. BN curves have embedding degree $12$), one needs to be careful for the choice of elliptic curves, and to change their sizes according to these new attacks. Using a more conservative but less efficient elliptic curves of embedding degree one would also be an alternative to implement pairing-based protocols. For the choice and right notion of types of pairings of embedding degree one we refer to the recent article of Chattarjee et al. \cite{Chattarjee2016}.
 
\subsection{Minimal Embedding Field Attacks} Hitt \cite{Lau07} observed that the minimal embedding degree $\mathbb{F}_p^{\text{ord}_N p}$ is not necessarily equal to the field $\mathbb{F}_q^k$, i.e. the extension can be defined over $\mathbb{F}_p$ instead of over $\mathbb{F}_q$. Hence, in this case the group $\mathbf{G}_T$ can be realized as a subgroup of much smaller field yielding to solve the DLP more efficiently in $\mathbf{G}_1$, $\mathbf{G}_2$ and $\mathbf{G}_T$. Note that this attack is only applicable for pairing-friendly curves defined over non-prime fields.

\subsection{Subgroup Attacks} Usually pairing functions are realized in such a way that two out of three groups $\mathbf{G}_1$, $\mathbf{G}_2$ and $\mathbf{G}_T$ are proper subgroups of larger composite order subgroups. This results in the so-called subgroup attacks if especially the underlying pairing implementation is not testing the group membership of the elements. Barreto et al. defined the concept of subgroup security and pointed out that most implementations of bilinear maps do not satisfy this notion \cite{BCMNPZ15}. They suggested new curve parameters using the known families of pairing-friendly elliptic curves achieving the subgroup security.

\section{Hard Problems Related to Pairing}\label{hard_problems}

There are plenty of pairing related computational and decisional problems. Their intractabilities form the basic security assumptions upon which pairing-based cryptographic protocols are designed. In this section, we only focus on the most general and frequently used hard problems.

\subsection{Pairing Inversion Problem}

A necessary straightforward security assumption is the one-wayness of the underlying pairing function $e$. The generalized pairing inversion problem (GPInv) asks to find $P$ $\in$ $\mathbf{G}_1$ and $Q \in \mathbf{G}_2$ such that $e(P, Q) =g$ for a given pairing function $e$ and a value $g$ $\in$ $\mathbf{G}_T$. This problem can be divided into two subproblems: 

\begin{itemize}
\item The fixed argument pairing inversion problem 1 (FAPI-1) is to find $Q \in \mathbf{G}_2$ such that $e(P, Q) =g$ for a given $P$ $\in$ $\mathbf{G}_1$ and $g$ $\in$ $\mathbf{G}_T$.

\item The fixed argument pairing inversion problem 2 (FAPI-2) is to find $P \in \mathbf{G}_1$ such that $e(P, Q) =g$ for a given $Q$ $\in$ $\mathbf{G}_2$ and $g$ $\in$ $\mathbf{G}_T$.
\end{itemize}

A simple observation shows that for each given pair $(P, g) \in \mathbf{G}_1 \times \mathbf{G}_T$ or $(Q, g) \in \mathbf{G}_1 \times \mathbf{G}_T$ both problems FAPI-$i$, $i=1,2,$ have a unique solution by non-degeneracy of $e$ and cyclicity of $\mathbf{G}_1, \mathbf{G}_2$ and $\mathbf{G}_T$. Note that in the explicit realization of pairings FAPI-$i$ (in terms of the size of the input $(P, g)$ or $(Q, g)$) can be solved \emph{at most} in subexponential time since there exists a subexponential DLP solver for $\mathbf{G}_T$ (in terms of the input size of $\mathbf{G}_1$, $\mathbf{G}_2$ and $\mathbf{G}_T$) by Theorem \ref{dlp-mov} and discussion in Section \ref{DLPattacks}. Again by the discussion in Section \ref{DLPattacks} and Theorem \ref{dlp-mov} it follows that FAPI-$i$ can even be solved in quasi-polynomial time since there exists a quasi-polynomial DLP solver for $\mathbf{G}_T$ for certain choices of $\mathbf{G}_1$, $\mathbf{G}_2$ and $\mathbf{G}_T$. For a detailed relation of the pairing inversion problems and the Diffie-Hellman type assumptions we refer to \cite{GHV11}.

Bilinear maps can be computed mainly in two stages. The first one is to compute the evaluation of a certain function at a certain divisor of the underlying elliptic curve $E$ by using Miller's algorithm \cite{BSS05}. The second stage is the \emph{final exponentiation}. For the details about the relationship between the individual steps (Miller inversion and inverting exponentiation) and the pairing inversion problem we also refer to \cite{GHV11}.

\subsection{Diffie-Hellman Related Problems}

Other most common pairing related problems are as follows:

\begin{definition}[Computational Bilinear Diffie-Hellman Problems \cite{BSS05}]

Let $e: \mathbf{G}_1 \times \mathbf{G}_2 \rightarrow \mathbf{G}_T$ be a non-degenerate bilinear pairing. Then
\begin{itemize}
\item The bilinear Diffie-Hellman problem 1 (BDH-1) asks to find $e(P, Q)^{ab}$ for given $P, aP, bP \in \mathbf{G}_1$, $Q \in \mathbf{G}_2$ and random $a,b$. 
\item The bilinear Diffie-Hellman problem 2 (BDH-2) asks to find $e(P, Q)^{ab}$ for given $P \in \mathbf{G}_1$, $aQ, bQ \in \mathbf{G}_2$ and random elements $a,b$.
\end{itemize}

\end{definition}

A frequently used variant of the decisional Diffie-Hellman problem in the Type-1 setting ($\mathbf{G}_1 = \mathbf{G}_2$) is given as follows:

\begin{definition}[Decisional Bilinear Diffie-Hellman Problem \cite{BSS05}]

Let $e: \mathbf{G}_1 \times \mathbf{G}_1 \rightarrow \mathbf{G}_T$ with a cyclic group $\mathbf{G}_1=<P>$ is given. Then
\begin{itemize}
\item The decisional bilinear Diffie-Hellman problem (DBDH) is to decide whether $h=e(P,P)^{abc}$ for given $P, aP, bP, cP$ $\in$ $\mathbf{G}_1$ with random elements $a,b,c$ and a random element $h\in\mathbf{G}_T$.
\end{itemize}

%$k$-Decisional Diffie-Hellman Inversion: Distinguish $P$, $yP$, $(y^2)P$, $\cdots$, $(y^k)P$, $e(P, P)^{1/y}$ from $P$, $yP$, $(y^2)P$, $\cdots$, $(y^k)P$, $e(P, P)^z$.

\end{definition}

It is clear that the decisional Diffie-Hellman problems including the pairing related ones are solvable in polynomial time when one has oracles solving the computational Diffie-Hellman problems. However, there are groups for which the classical decisional Diffie-Hellman problem is easy while the classical computational Diffie-Hellman problem is believed to be hard. In particular, a gap Diffi-Helman group has a distinguishability oracle for which solving the computational problem is hard \cite{BLS01,JN03}. In the Type-1 pairing setting the Gap Diffie-Hellman Problem is formally defined as follows:

\begin{definition}[Gap Diffie-Hellman Problem \cite{BLS01,JN03}]
Given groups $\mathbf{G}_1$ and $\mathbf{G}_2$ of prime order $q$, a bilinear map $e: \mathbf{G}_1 \times \mathbf{G}_1 \rightarrow \mathbf{G}_T$ and a generator $P$ of $\mathbf{G}_1$. The Gap Diffie-Hellman Problem (Gap DH) asks to compute $abP$ for given instance $(P, aP, bP)$ of the CDH problem and a DDH oracle.

\end{definition}

\begin{definition}[Co-Assumptions \cite{BGLS03}]

\begin{itemize}
\item The Computational Co-Diffie-Hellman Problem asks to compute $aQ$ for given $P$, $aP$ $\in$ $\mathbf{G}_1$ and $Q$ $\in$ $\mathbf{G}_2$ for a random element $a$.

\item The Decisional Co-Diffie-Hellman Problem asks to decide whether $aQ=R$ for given $P$, $aP$ $\in$ $\mathbf{G}_1$ and $Q$, $R$ $\in$ $\mathbf{G}_2$, for a random element $a$.

\end{itemize}

Assume additionally that $\mathbf{G}_1$ $\neq$ $\mathbf{G}_2$. Then,

\begin{itemize}
\item The Computational Co-Bilinear Diffie-Hellman Problem asks to compute $e(P,Q)^{abc}$ $\in$ $\mathbf{G}_1$ for given $(P, aP, bP)$ $\in$ $\mathbf{G}_1^3$ and $(Q,aQ,cQ)$ $\in$ $\mathbf{G}_2^3$ for random elements $a,b$ and $c$.

\item The Decisional Co-Bilinear Diffie-Hellman Problem asks to distinguish $P,aP,bP,Q, e(P,Q)^{ab}$ from $P, aP, bP, Q$, $e(P,Q)^{z}$ for random elements $a,b$ and $z$.

\end{itemize}
\end{definition}

It is trivial to see that the Decisional Co-Diffie-Hellman problem is easy to solve if we have an efficiently computable bilinear map. 

The relationship between the CDH and FAPI-1 and FAPI-2 problems is given by the following theorem of Galbraith et al. \cite{GHV11} whose proof follows for a CDH instance $(P,aP,bP)$ easily from first calling the FAPI-1 oracle with the inputs $(P, e(aP,Q))$ to obtain $aQ$ for a random element $Q$ and calling secondly the FAPI-2 oracle $(Q, e(bP,aQ))$:

\begin{theorem}
 Let $e: \mathbf{G}_1 \times \mathbf{G}_2 \rightarrow \mathbf{G}_T$ be a non-degenerate bilinear pairing on cyclic groups of prime order $r$. Suppose one can solve FAPI-1 and FAPI-2 in polynomial time. Then one can solve the computational Diffie-Hellman problem in $\mathbf{G}_1$, $\mathbf{G}_2$ and $\mathbf{G}_T$ in polynomial time.
\end{theorem}

Similar to the above argumentation the following result is also proved in \cite{GHV11}:

\begin{theorem}

Let notation be as above. If one can solve FAPI-1 (resp. FAPI-2) in polynomial time then one can compute all non-trivial group homomorphisms $\phi_2: \mathbf{G}_2 \rightarrow \mathbf{G}_1$ (resp. $\psi_2: \mathbf{G}_2 \rightarrow \mathbf{G}_1$) in polynomial time.

\end{theorem}

We continue with an assumption which is frequently used in the design of pairing-based protocols: 

\begin{definition}[The external Diffie-Hellman (XDH) assumption \cite{BBS04}]
Let the CDH be intractable in both $\mathbf{G}_1$ and $\mathbf{G}_2$. The external Diffie-Hellman assumption (XDH) states that the DDH is also intractable in $\mathbf{G}_1$. If the DDH is also intractable in $\mathbf{G}_2$ we have the symmetric external Diffie-Hellman assumption (SXDH).
\end{definition}

\begin{remark}
It is easy to see that the GDH problem is only realizable with the Type-1 pairing, and the strict XDH assumption (i.e. if SXDH does not hold) corresponds exactly to the Type-2 setting. Furthermore, the SXDH assumption is only realizable in the Type-3 setting. 
\end{remark} 

There are also several cryptographic protocols whose security relies on other pairing related problems with auxiliary inputs:

\begin{definition}[Pairing problems with auxiliary inputs \cite{Cheon:2016:NAD}]
Let the elements $g,g^{\alpha}, \cdots , g^{\alpha^d}$ in $\mathbf{G}_1$ (resp. $\mathbf{G}_2$) be given with a random element $\alpha$. Then, the DLP with auxiliary inputs (DLPwAI) is to compute $\alpha$. Solving the DLPwAI implies the solution of many pairing-based problem assumptions. These are called pairing problems with auxiliary inputs. 

These include the Weak Diffie-Hellman (wDH) Problem, the Strong Diffie-Hellman (sDH) Problem, the Bilinear Diffie-Hellman Inversion (BDHI) Problem and the Bilinear Diffie-Hellman Exponent (BDHE) Problem.
\end{definition}

\begin{remark}
The generalized DLP with auxiliary inputs problem (GDLPwAI) is to compute a randomly chosen $\alpha$ if $g,g^{\alpha^{e_1}}, \cdots , g^{\alpha^{e_d}}$ in $\mathbf{G}_1$ (resp. $\mathbf{G}_2$) are given and $K:=\{e_1,\cdots ,e_d\}$ is a multiplicatively closed subset of $\mathbb{Z}_{r-1}^{\times}$ \cite{CKS13}.

We note that some generalized versions of the Weak Diffie-Hellman (wDH) Problem, the Strong Diffie-Hellman (sDH) Problem, the Bilinear Diffie-Hellman Inversion(BDHI) Problem and the Bilinear Diffie-Hellman Exponent (BDHE) Problem can also be vulnerable if the GDLPwAI is solved. 
\end{remark}

\section{Security or Efficiency Issues of Recent Papers}\label{issues}

In this section, we revisit a collection of recently proposed research papers in order to illustrate the incorrect use of the pairing-based primitives. 

\begin{itemize}
\item In \cite{HCCS15}, the authors propose a batch verification mechanism which aims to verify multiple digital signatures at a time less than the total individual verification time. The authors prove the security of their scheme under the collusion attack assumption using the Type-1 setting with supersingular elliptic curves \cite[pp.2526]{HCCS15} having security issues as discussed in Section \ref{DLPattacks}. We note also that this issue still applies most recent papers in cryptography journals such as \cite{Hofheinz2015,Libert2015,Okamoto2015,Lee2016}.

\item In \cite{WSXF15}, the authors present a new Type-1 pairing-based multi-receiver encryption scheme and authenticated key establishment protocol for vehicular ad-hoc network (VANET). Its security analysis relies on the system of \cite{THC12} which is based on the underlying Gap DH, hence realizable only in the Type-1 setting. Their example with a 512 bit supersingular elliptic curve with embedding degree 2 is too inefficient since the same security level can be guaranteed in the asymmetric setting for instance with a Barreto-Naehrig curve of 160 bits \cite{FST10}.

\item In \cite{LZJ15}, the authors proposed an authenticated encryption system using the Type-1 setting aiming to accomplish confidentiality and authenticity simultaneously. This scheme is applied to email system as a practical example. Unfortunately, the proposed example does not have any complexity advantage over the current system. The scheme is also too inefficient because of the use of the Type-1 setting as in the previously mentioned schemes.

\item In \cite{GCGF15} the authors propose a privacy-preserving scheme for incentive-based demand response in the smart grid. The smart grid technology basically uses the information and communication technologies aiming to enhance the efficiency, reliability, sustainability of the generation, transmission, distribution, and consumption of electricity. They used a Type-1 pairing although the security of the scheme relies on the existence of an efficiently computable homomorphism in the Type-2 setting as stated in \cite{GCGF15} using the unforgeability of BBS+ signature \cite{BBS04}. Therefore, the scheme has to be modified into a more efficient Type-3 setting in the light of \cite{CM11}.

\item Unlike wired networks, mobile ad-hoc networks (MANETs) are more vulnerable to some attacks bringing new security challenges (e.g., limited resources, open peer-to-peer network, dynamic network topology, lack of a trusted centralized authority). Therefore, designing and implementing more efficient cryptographic algorithms, key management, secure neighbor detection, and routing protocol are some of the active research areas. Certificate-less (mostly pairing-based) public key based solutions are known to be one of the best candidates. But after surveying the pairing-based MANETs, we again realize the incorrect use of the Type-1 setting and counting the pairing operation as a black-box, see for instance \cite{GD15}.

\item In \cite{Chu:2014:KCS:2574228.2574550}, the authors proposed a solution for scalable data sharing in cloud storage using key-aggregate cryptosystems using the Type-1 setting. A subsequent paper \cite{Liu2014} deals with a cloud data sharing scheme utilizing supported keyword search. It also suffers from the use of the Type-1 setting. For instance, in the latter paper, the security of the scheme relies on the intractability of the computational Diffie-Hellman problem which is no longer secure for supersingular elliptic curves over small characteristic finite fields as discussed in Section \ref{DLPattacks}.

\item The authors in \cite{Liu:2015:EIV:2778378.2778558} proposed a mechanism using the Type-1 setting for data integrity verification for the Internet of Things (IoT) applications, where the integrity of an outsourced data is the most crucial security property. The authors did not unfortunately modify the original BLS idea into the Type-3 setting. 

\item In \cite{HJG14} the authors proposed secure data transmission mechanism for cluster-based wireless sensor networks using the Type-1 setting. Their analysis in \cite[pp. 758]{HJG14}, however, cannot be realized in this setting. Therefore, the proposed scheme is much less efficient than the author's quantitative calculation.

\item The author proposes in \cite{Wan15} a remote data integrity checking model in multi-cloud platforms. The scheme uses the Gap DH assumptions like most of their counterparts (public auditing schemes) and hence uses the Type-1 setting. In the simulation of the scheme the author argued to use 160 bit elliptic curve for the underlying bilinear map. It is impossible to obtain a secure mechanism using such a small order elliptic curve (either insecure for curves over a field of characteristic $3$ or insecure since the embedding degree is $2$ over large fields due to the discussions in Section \ref{DLPattacks}.)

\item In \cite{CN03}, Coron and Naccache proved that the Co-Diffie Hellmann problem and the $k$-element aggregate extraction problem are equivalent with the assumption that there exists an efficiently computable homomorphism $\phi:\ \mathbf{G}_2\rightarrow \mathbf{G}_1$. Recently, in \cite{XZ16} Xie und Zhang proposed a secure incentive scheme for delay tolerant networks under the $k$-element aggregate extraction problem using the Type-1 setting. However, it is impossible to convert their scheme into the Type-2 setting. The reason is that their scheme uses the existence of an efficient and secure hashing to point map into $\mathbf{G}_2$, which is in fact not realizable in the Type-2 setting as discussed in Section \ref{prel}.

\item Kundu and Bertino \cite{KB13} proposed an authentication mechanism using the Type-1 bilinear maps under the $k$-element aggregate extraction problem in order to achieve confidentiality-preserving authentication of trees and graphs. As in the above mechanism, an asymmetric modification of the scheme can only be realized in the Type-2 setting since the author uses the original BGLS \cite{BGLS03} aggregate signatures. However, this cannot be realized either, since one requires efficient and secure hashing to a point map into $\mathbf{G}_2$.

\end{itemize}

\section{Recipe for Designers}

The following conditions have to be taken into consideration by designing cryptosystems using bilinear maps:

\begin{itemize}

\item \textbf{Use the Type-3 Setting:} Although there are automated tools converting protocols using the Type-1 bilinear maps into protocols using the Type-3 bilinear maps \cite{AGOT14,AGH15} and a general framework converting the Type-2 schemes into the Type-3 schemes \cite{CM11}, it is always better to design cryptosystems using directly the Type-3 pairing to achieve the best security level and the most efficient protocols. However, one must also be careful about the new complexity bounds on the DLP if the chosen embedding degree is $6$ or $12$ \cite{Kim2016}.

\item \textbf{Choose the Best Pairing Function:} One should use efficient computation in pairing-friendly groups. In other words use optimal pairing \cite{FHess:2008a} or its efficient variants as much as possible. The pairing function should be specified in order to obtain the best efficiency security trade off.

\begin{table*}[htpb]
 \caption{Revised version of the comparison of different pairing types \cite[Table 1]{GPS08}. A checkmark $\checkmark $ denotes that the pairing type satisfies the property and \xmark \space denotes that it fails to satisfy the property. The \# of * measures the efficiency of the underlying pairing type (*** denotes the most efficient choice). Note that $p$ denotes the characteristic of the field over which the curve is defined (i.e. over $\mathbb{F}_q$ with $q = p^n$).}\label{comparison}

\begin{center}
\begin{tabular}{|c|c|c|c|c|c|c|}
    \hline
     Type& Hash to & Short &Homomorphism& Poly time & Security & Efficiency\\ 
		& $\mathbf{G}_2$& $\mathbf{G}_1$ & & generation & & \\ \hline
		
   1 ($p$=2 or 3)& $\checkmark$ & \verb|| \xmark &$\checkmark$&\verb|| \xmark &\red{\verb|| \xmark}& **\\ \hline
    $1 (\text{large } p)$ & $\checkmark$ &\verb|| \xmark & $\checkmark$ & $\checkmark$&$\checkmark$&\red{*} \\ \hline
		2&\verb|| \xmark&$\checkmark$&$\checkmark$&$\checkmark$&$\checkmark$&** \\ \hline
		3&$\checkmark$&$\checkmark$&\verb|| \xmark&$\checkmark$&$\checkmark$&*** \\ \hline
    \end{tabular}
\end{center}
\end{table*}

\item \textbf{Begin with a Correct Set-Up:} Implementation details should be given more concretely (\textit{what is the desired security level?}) and always together with its usability (\textit{which pairing Type is used?}) and practical aspects (\textit{what is the computation and communication overhead?) and realizability aspects (which pairing function has to be used?}.

\item \textbf{Use Realizable Security Assumptions:} It is crucial to avoid unrealizable security assumptions. Furthermore, the assumptions about the security level should be carefully stated using concrete constraints. For a typical example, one may believe that ``it is always easy to generate efficiently suitable system parameters for pairing-based cryptosystems'' which is clearly wrong as outlined above. See Table \ref{recommendedsecuritylevels} for the concrete realizability constraints.

\item \textbf{Do Not Use Extensions of Binary and Ternary Curves:} One should be careful about more destructive security issues resulting from attacks on the DLP over fields of small characteristics (following the lines of Section \ref{DLPattacks}).

\item \textbf{Avoid the Explicit Homomorphisms:} A possible wrong use of the asymmetric setting with the assumption of the existence of efficiently computable homomorphisms leads to unrealizability and/or security and privacy leakage. We note that in some applications there is a tendency to use the asymmetric setting incorrectly with the assumption of the existence of efficiently computable homomorphisms in both directions (both from $\mathbf{G}_1$ to $\mathbf{G}_2$ and $\mathbf{G}_2$ to $\mathbf{G}_1$). See Table \ref{recommendedsecuritylevels} for the concrete realizability constraints.

\item \textbf{Use the Hashing to Point only in the Type-3 Setting:} Pairing-based cryptographic protocols may require the following underlying assumptions simultaneously: (1) secure and efficient hashing into group elements (2) efficient homomorphism from $\mathbf{G}_2$ to $\mathbf{G}_1$. This requirement can be vital in order to prove the security of the underlying protocol or to design comparably more efficient mechanisms. However, since both requirements cannot be realized simultaneously in practice, the design criteria should be checked carefully in order to ensure the claimed security and efficiency while achieving a realizable mechanism. See Table \ref{recommendedsecuritylevels} for the concrete realizability constraints.

\item \textbf{Test Group Membership or Use Subgroup Secure Curves:} In order to undermine the implementation attacks (for instance, failing to test the group membership) subgroup security has to be guaranteed in the realization of pairing-based protocols by generating subgroup secure pairing-friendly elliptic curves following the lines of \cite{BCMNPZ15}.

\item \textbf{Do Not Use Curves over Extension Fields, Use Prime Fields:} In order to estimate the desired level of security precisely, special caution on the realizability of the minimal embedding field attack has to be taken if the underlying elliptic curves are defined over extension fields. In particular, the equality $\text{ord}_N(p)=mk$ needs to be hold, where $q=p^m$, $\mathbf{G}_1$ is a subgroup of $N-$th torsion subgroup of the underlying elliptic curve over $\mathbb{F}_q$ and $k$ is the embedding degree.  

\item \textbf{Take the Correct Decision on the Key Sizes:} In order to use composite embedding degrees, the key sizes of the underlying elliptic curves need to be updated by using the complexity results of recent attacks on the medium size finite fields of composite degrees \cite{Kim2016,Sarkar2016,Jeong2016}. The complexity estimates of these new attacks suggest that doubling the sizes of the underlying elliptic curves or using elliptic curves of embedding degree one are already safe at the cost of additional computational overhead.

\item \textbf{Be Careful about the Auxiliary Inputs:} In order to avoid possible attack scenarios caused by solving the discrete logarithm with auxiliary inputs (DLPwAI) \cite{Cheon:2016:NAD} and its generalizations (GDLwAI) \cite{CKS13} special caution has to be taken by the choice of the underlying elliptic curves and the orders of $\mathbf{G}_1$ and $\mathbf{G}_2$. Especially, for the order $r$ of $\mathbf{G}_1$ and $\mathbf{G}_2$ the values $r-1$ and $r+1$ should have no small divisors. Moreover, auxiliary exponents should not be closed with respect to the multiplication. This is important if one needs the security assumptions like the Weak Diffie-Hellman (wDH) Problem, the Strong Diffie-Hellman (sDH) Problem, the Bilinear Diffie-Hellman Inversion(BDHI) Problem and the Bilinear Diffie-Hellman Exponent (BDHE) for the design of the pairing-based protocols.

\end{itemize}

\section{Conclusion}
In this paper, we aim to highlight once again the wrong usage of bilinear maps in the recent research papers which unfortunately leads to security, realizability and/or efficiency issues. Furthermore, with the practicality and advantages of pairing-based technologies researchers should focus on the correctness and the mathematical details instead of using them as a ``black-box''. Moreover, the National Institute of Standards and Technology (NIST) and IEEE have been actively working on the correct versions of pairing-based cryptography to bring them to the state-of-the-art advancements, but the current versions are vulnerable to the recent attacks \cite{IEEE13, Nist15}.

% This solves the open problem formulated in \cite{CLMTL12}.

%\bibliographystyle{unsrt} % abbrv ?
\bibliographystyle{abbrv} % abbrv ?
\bibliography{Pairingsurvey}

\appendix

\section*{Appendix}\label{appendix}

\section{Application Areas of Pairings}

Pairing-based cryptography is being considered for alternative constructions in many areas of cryptographic research. It is an active research area for deploying novel security and privacy mechanisms, e.g., \cite{EM14,YOK--JZCK14, Fre14, HCCS15,CWS15,NR15,ZQWSDH15, WSXF15,LZJ15,ZDMSTR15, BCMNPZ15,CM15,GCGF15, Tsa15, GD15, ANBYSTYNH15}. These include the following applications:

\subsubsection{Identity based encryption (IBE) \cite{BF01}:} It is a special type of public-key encryption in which a publicly known identifier is used as a public key. More concretely, a trusted third party first generates its public/private key pair which is called ``master'' public key and ``master'' private key. Next, a user's public key is replaced with an identity (e.g., an email, an address, a photo, a phone number, a post address) and his/her private key is computed based on the identity and master private key. IBE allows the user to send an encrypted message to another user using his/her identity as a public key and the user decrypts it with the corresponding public key. IBE based schemes do not require public-key generation and distribution as it exists in the conventional public key systems, which significantly reduce/eliminate the cost and complexity of generating and managing users' certificates (i.e., a public key infrastructure). It has further an interesting property that private keys need not to be generated before sending an encrypted message.

\subsubsection{Hierarchical identity-based Encryption (HIBE) \cite{Boneh2005}:} It allows the private key generator to delegate its computation to the lower-level private key generators. Furthermore, anonymous HIBE is an extension of IBE which hides not only the message itself but also the identity of the users. Anonymous HIBE solutions can be applied to anonymous communication systems and public key encryption systems with a keyword searching mechanism.

\subsubsection{Functional (or Attribute based) Encryption \cite{Lewko2010,Lewko2011}:} It uses pairings to generate decryption keys which allows a user possessing an encrypted data {\sf Enc}$(x)$ to compute $f(x)$ of the data for an arbitrary function $f$. 

\subsubsection{IBE with threshold decryption \cite{BZ04}:} The master key of the trusted third party of a standard IBE system can be distributed in a $(k,n)$ fashion among $n$ different independent authorities, where at least $k$ of them must cooperate and collude to perform decryption (using conventional techniques of threshold cryptography like Shamir secret sharing schemes).

\subsubsection{Searchable encryption \cite{Fang2013}:} It allows a user to compute whether a given keyword exists in an encrypted message without giving away any information about the message itself. In practice, it is possible to search any query on an encrypted database without decryption (e.g., patient medical records, biometric data, personal data, corporate data, intellectual property).

\subsubsection{Signatures \cite{BSS05,SW08}:} Digital signatures is an important primitive which ensures authentication, integrity of a message, and non-repudiation. Apart from conventional signature schemes (based on RSA or ECC) pairing/ID based signatures are constructed because of some nice structural properties like homomorphic linear authenticators where the authenticators can be aggregated into only one tag, which significantly reduces the communication and computational complexity. Other types of pairing-based signature schemes include short signatures (also without random oracles), blind signatures (where a user obtains a signature from a signer while the signer does not learn any information about the message being signed), identity based signatures (also including ID-based blind signatures, hierarchical ID-based signatures, ring signatures), chameleon signatures (non-repudiable and non-transferrable), aggregate signatures (which allows multiple signatures to be aggregated into one compact signature), ring signatures (where any group member can sign a message without learning any information about the signed message), group signatures (which is similar to ring signatures except that a ``group manager'' can detect which group member indeed signed a message), threshold signatures (a valid signature can be computed only if at least $t$ signers cooperate), authentication-tree based signatures without random oracles.

\subsubsection{New security requirements for cloud \& IoT security:} Privacy enhancing techniques (like privacy-preserving auctions, anonymous credentials, or privacy-friendly aggregation for the smart grid), proofs of retrievability of data for cloud storage systems \cite{SW08}, internet of things (IoT) \cite{LX13}, e-health systems and wearable technologies \cite{LHBC12}.

\subsubsection{Other applications:} Last but not least, there are also various ID based mechanisms including authentication \cite{Li2009,RFCAuth12}, identity based key-agreement \cite{Chen2007, Wang2013}, signcryption (which is a public key authenticated encryption, i.e. including both signing and encrypting operations simultaneously), and identity based Chameleon hashes \cite{Ateniese2004} (which are collision resistant functions with a trapdoor for finding collisions).

\end{document}